\documentclass[aps,pre,twocolumn,superscriptaddress,showpacs,showkeys,floatfix]{revtex4-1}
\usepackage{amsmath}
\usepackage{amsfonts}
\usepackage{graphicx} 
\usepackage{amssymb} 
\usepackage{grffile}
\usepackage{color}

\newcommand{\Zgpf}{\mathcal{Z}}

\newcommand{\beq}{\begin{equation} }
\newcommand{\eeq}{\end{equation}}
\newcommand{\tildet}{{\tilde{t}}}
\def \beqa #1\eeqa{\begin{align}#1\end{align}}

\begin{document}
\title{Urn model for products' shares in international trade}
\author{Matthieu Barbier}
\affiliation{Centre for Biodiversity Theory and Modelling, Theoretical and Experimental Ecology Station, CNRS, Moulis 09200 , France}
\author{D.-S.  Lee}
\affiliation{Department of Physics, Inha University, Incheon 22212, Korea}
\email{deoksun.lee@inha.ac.kr}

\date{\today}

\begin{abstract}
International trade fluxes evolve as countries revise their portfolios of trade products towards economic development.  Accordingly products' shares in international trade vary with time, reflecting the transfer of capital  between distinct industrial sectors. Here we analyze the share of hundreds of product categories in world trade for four decades  and find  a scaling law obeyed by the annual variation of product share, which informs us of how capital flows and interacts over the product space.  A model of stochastic transfer of capital between products based on the observed scaling relation is proposed and shown to reproduce exactly the empirical share distribution. The model allows analytic solutions as well as numerical simulations, which predict a pseudo-condensation of capital  onto few product categories and when it will occur.   At the individual level, our model finds certain products unpredictable,  the excess or deficient growth of which  with respect to the model prediction is shown to be correlated with the nature of goods.
\end{abstract}

\maketitle 

\section{Introduction}
\label{sec:intro}

Finite and uneven distribution of resources and capabilities for production lead a huge volume of products and capital to be exchanged across countries. The organization of such international trade has long been studied to elucidate the impact of the international relations, geography and sociocultural factors on trade fluxes~\cite{tinbergen1962analysis}, as represented by e.g., the gravity model~\cite{anderson2010gravity,barigozzi2010multinetwork}, as well as the topology of  the trade network of countries~\cite{PhysRevE.68.015101,PhysRevLett.93.188701}.

The fluctuation of trade volumes of  various products also carries  valuable information on human economic activities.  As a country's portfolio of exports is crucial for both its immediate success in the global market   and long-term development~\cite{hidalgo2007product,hidalgo2009building,hausmann2011network,tacchella2012new}, political and economic agents often shift their investment strategy from one product to another, which affect the market share of related products.  Aggregated together, a huge number of such strategy shifts across countries, time, and products cause fluctuations of products' shares, which can provide a clue for understanding a principle of  capital management towards best benefiting  from producing and exporting selected goods.

Here we report our finding of a scaling behavior of the annual variation of trade products' shares in international trade. It is shown to imply that  capital invested for various product categories, identified here with their shares, tends to move to and from popular products, the probability of which is quadratically proportional to their current shares.  A stochastic model of such biquadratic  transfer of discretized capital among products is proposed, which reproduces the empirical distribution and evolution of product share, confirming that the shifts of investment strategy are made mostly referring to the  current shares of products. Random transfer models have been similarly proposed to explain the distribution of wealth among individuals or countries~\cite{yakovenko2009colloquium,bouchaud2000wealth}. The functional form of the transfer rate determines whether the system is either in the fluid phase or the condensate phase with the biquadratic one at the boundary. Analytic solutions and simulation results reveal the possibility of condensation of capital onto few key products and predict its time scale.  Therefore the studied model can be used as a framework helping understand and analyze the overwhelming complexity of international trade. 

\section{Distribution and annual variation of product share}
\label{sec:empirical}

\begin{figure}
\centering
\includegraphics[width=\linewidth]{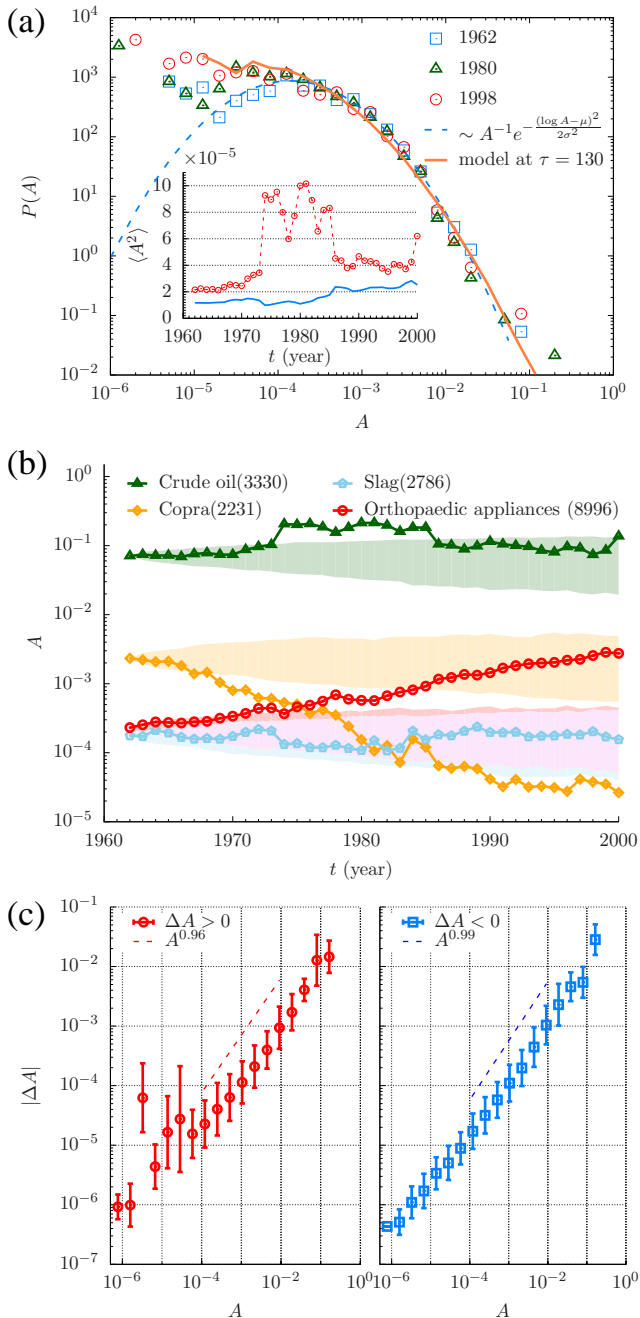}
\caption{
Product share in international trade.
(a) Probability density function of product share  $A$. Empirical data (symbols) for year 1962 are fitted by a lognormal function $P(A)\sim A^{-1} \exp\left[-{(\log A - \mu)^2 \over 2\sigma^2 }\right]$with $\mu=-7.1$ and $\sigma=1.3$ (dashed line)~\cite{doi:10.1137/070710111,Alstott:2014aa}. The solid line is taken from model simulations 
at the rescaled time $\tau=130$ with $N=508$ products and the particle density $\rho=400$ detailed in Sec.~\ref{sec:modelb2}.
Inset: The second moment $\langle A^2\rangle = N^{-1} \sum_p A_p(t)^2$ versus $t$. The solid line is obtained by excluding  crude oil (3330).
(b) Evolution of individual products' share, denoted by name and 4-digit SITC code. While the share of  SITC 3330 and that of 2786 stay mostly contained in the central 80\% of simulated trajectories (shaded areas), 2231 and 8996 show large deviations from model predictions detailed in Sec.~\ref{sec:individual}. 
(c) Dependence of the annual variation of share  $\Delta A_p(t)=A_p(t+1) - A_p(t)$ on $A_p(t)$ in case of both (left) $\Delta A_p>0$ and (right) $\Delta A<0$. 
}
\label{fig:empirical}
\end{figure}

We analyze the NBER-UN dataset \cite{feenstra2005world}, which contains trade volumes of $N=508$ product types based on the Standard International Trade Classification (SITC Rev. 2)  and consistently reported over the period 1962-2000.
A single datum  $V_{c,c',p}(t)$ is the amount of trade (in nominal dollars) from country $c$ to country $c'$ in product category $p$ in the year $t$.  
We investigate product share
\begin{equation}
A_p(t) = \dfrac{\sum_{c,c'} V_{c,c',p}(t)}{\sum_{c,c',p'} V_{c,c',p'}(t)},
\label{eq:Apt}
\end{equation}
which is the fraction of the total amount of money involved in trade worldwide that is exchanged with product $p$, and thus represents the popularity of this product in the global economy.  
While the total trade volume $V_{\rm all}(t)=\sum_{c,c',p} V_{c,c',p}(t)$ grows exponentially, we are here interested in the investment-strategy shifts among various product categories, which are entirely contained in the dynamics of $A_p(t)$. 

Remarkably a small number of products occupy a large fraction of total trade~\cite{mantegna1995scaling}. Crude oil (SITC 3330) is the category with the largest share, retaining close to $10\%$ of all trade for the whole period. By contrast,  the share for e.g. the category of slag and similar waste (SITC 2786) never exceeds $10^{-3}$.  The general shape of the probability density function of product share $P_t(A) =\sum_p \delta(A_p(t)-A)$ is close to a lognormal function $P_t(A) \sim  A^{-1}\exp\left[-{(\log A-\mu)^2\over 2\sigma^2}   \right]$ as shown in Fig.~\ref{fig:empirical} (a)  or a stretched-exponential function $P_t (A) \sim A^{\beta -1} \exp [-\lambda A^\beta]$~\cite{doi:10.1137/070710111,Alstott:2014aa}.

Many products' share exhibits significant variation over time, as in the case of orthopedic appliances (8996) and copra (2231) which  display steady growth and decline  respectively over orders of magnitude in their shares [Fig.~\ref{fig:empirical} (b)].
We find that the annual variation of the share of an individual product category $\Delta A_p(t) = A_p(t+1)-A_p(t)$, when averaged over  gains ($\Delta A_p>0$) or losses ($\Delta A_p<0$), is found to be proportional to  the current share as 
\begin{equation}
|\Delta A_p (t)| \simeq c A_p(t)
\label{eq:DApAp}
\end{equation}
with $c\simeq 0.1$ [Fig.~\ref{fig:empirical} (c)]. For $A$ so small as $A\lesssim 10^{-5}$, the gain of product share shows fluctuations. 
The  linear scaling in Eq.~(\ref{eq:DApAp})  appears also in other economic time series~\cite{gibrat1931inegalites,mantegna1995scaling}. 
Considering $A_p(t)$ as approximating the time average $\langle A_p\rangle$ of $p$ in the period between $t$ and $t+1$ and $\Delta A_p(t)$ as  the standard deviation $\sigma_p$  in the same period, we find Eq.~(\ref{eq:DApAp}) represent the fluctuation scaling $\sigma_p \sim \langle A_p\rangle^\alpha$ with $\alpha=1$, which has been investigated for diverse complex systems~\cite{menezes04a,eisler08,PhysRevE.83.066115} under such names as Gibrat's law~\cite{gibrat1931inegalites} or Taylor's law~\cite{TAYLOR:1961aa}.  Though simple, Eq.~(\ref{eq:DApAp}) has far-reaching implications for the underlying dynamics of capital invested for product categories, which we address in the next sections.

\section{Urn model}
\label{sec:model}

Differentiated evolution of individual product share arise from a huge number of microscopic changes of investment made by companies and countries,  which can be modeled as  independent  trajectories subject to non-gaussian stochasticity \cite{mantegna1995scaling} or as emerging from exchange between individuals~\cite{yakovenko2009colloquium,bouchaud2000wealth}. We take the latter approach to understand the origin of the empirical features of product share presented in Sec.~\ref{sec:empirical}. 

\begin{figure}
\centering
\includegraphics[width=\columnwidth]{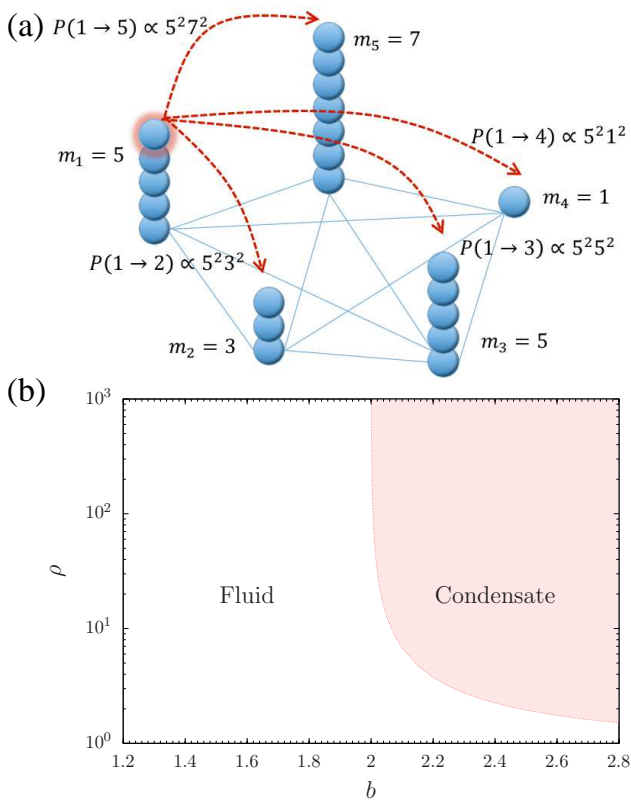}
\caption{Particle transfer and phase diagram of the urn model.
(a) Stochastic transfer of a particle (unit of share) for  $N=5$ sites (products) with rate 
 proportional to the square of the number of particles at source and destination  sites as $m_p^2 m_q^2$.
(b) Phase diagram  in the ($b,\rho$) plane with particle density $\rho=M/N$ and the scaling exponent $b$ of Eq.~(\ref{eq:u}). The boundary between fluid and condensate phase is given by Eq.~(\ref{eq:rhoc}). 
}
\label{fig:model}
\end{figure}

As the first step of modeling the kinetics of product share, we discretize the product share by introducing a unit share $a$, representing the amount available for a single strategic change. Then we find the continuous share $A_p$  discretized  to  $m_p = A_p/a$ particles, resulting in  a total of $M=\sum_{p=1}^N m_p = 1/a$ particles distributed over $N$ urns or sites (products). Transferring a particle from a site to another represents the flow of capital between them caused by the change of investment strategy of microscopic agents on a short time scale. We assume that those {\it share} particles hop from a site to another stochastically and independently. Such a particle-hopping model can be classified as  the urn model~\cite{godreche2002nonequilibrium,0305-4470-38-19-R01} in which a site $p$ sends one of its particles to another site $q$ with rate $u_{pq}$ varying with both the departure and destination site [Fig.~\ref{fig:model} (a)]. 

\subsection{Biquadratic transfer rate}

The transfer rate $u_{pq}$ can be determined to satisfy Eq.~(\ref{eq:DApAp}). A sequence of  $\mathcal{N}$ particle jumps in or out of a site $p$ sums up to increase or decrease the number of particles at $p$ by $|\Delta m_p| \sim \sqrt{\mathcal{N}}$ as in  random walks~\cite{hughesbook}. Therefore the linear scaling in Eq.~(\ref{eq:DApAp}) can be reproduced if a site $p$ is selected  $\mathcal{N} \propto m_p^2$ times per unit time.  
This reasoning leads us to the transfer rate $u_{pq}$ being {\it biquadratic} in the numbers of particles at $p$ and $q$. That is, $b=2$ in 
\begin{equation}
u_{pq}=u(m_p,m_q) = \dfrac{(m_p m_q)^b }{ \sum_{\ell} m_\ell^b}.
\label{eq:u}
\end{equation} 
If $b=0$, sites are uniformly selected,  the number of particles at a site follows the Boltzmann distribution in the stationary state $p_\infty(m)=\rho^{-1} e^{-m/\rho}$ with the particle density $\rho\equiv M/N$~\cite{yakovenko2009colloquium}. If $b\neq 0$, sites are given unequal chances for particle hopping.  If we consider particle hopping as occurring due to attraction between particles and assume that $u_{pq}$ is proportional to the sum of those attractions between all pairs of particles at site $p$ and $q$, we find that individual particles attract one another with equal strength for $b=1$. The quadratic scaling $b=2$, reproducing the linear scaling in Eq.~(\ref{eq:DApAp}), implies inhomogeneous interaction between particles; the interaction  between a particle at site $p$ and another at $q$ is  proportional to $m_p m_q$. It should be mentioned that a site having only one particle is not allowed to send a particle elsewhere, i.e., $u_{pq}=0$ if $m_p=1$, since our study is restricted to the  products maintaining non-zero share in the studied period.

\subsection{Stationary-state distribution and phase diagram}
\label{sec:pd}
To see how particles are distributed under Eq.~(\ref{eq:u}), let us consider  the evolution of the probability of finding a particle configuration $\{m\}=\{m_p|p=1,2,\ldots, N\}$ with time $\tildet$~\cite{0305-4470-38-19-R01}:
\begin{align}
& {\partial  p_{\tildet} (\{m\})  \over \partial\tildet} = 
\sum_{p<q} 
\left[u(m_p+1, m_q-1) \, p_\tildet(\{m\}_{pq}^{(+-)}) \right. \nonumber\\
& - u(m_q, m_p) \, p_\tildet(\{m\}) +  u(m_q+1, m_p-1)  \, p_\tildet(\{m\}_{pq}^{(-+)}) \nonumber\\
 &\left.  - u (m_p, m_q) \, p_\tildet(\{m\}) \right],
\end{align}
where $\{m\}_{pq}^{(\pm, \mp)}$ is identical to $\{m\}$ except at sites $p$ and $q$ such that $\{m\}_{pq}^{(\pm, \mp)}= \{m_1, m_2, \ldots, m_{p-1}, m_p\pm 1, m_{p+1}, \ldots, m_{q-1}, m_q\mp 1, m_{q+1}, \ldots\}$.  In the stationary state $(\tildet\to\infty)$, the detailed balance condition  $u(m_p+1, m_q-1) \, p_\infty(\{m\}_{pq}^{(+-)})  = u(m_q, m_p) \, p_\infty(\{m\})$ (and equivalently $u(m_q+1, m_p-1)  \, p_\infty(\{m\}_{pq}^{(-+)})  = u(m_p, m_q) \, p_\infty(\{m\})$) is satisfied if  $u(m_p, m_q)$ is  the multiplication  of a function of $m_p$ and a function of $m_q$~\cite{0305-4470-38-19-R01}, which is approximately true  in the stationary state of our model; the denominator of Eq.~(\ref{eq:u}) saturates in the long time limit as confirmed in Fig.~\ref{fig:density} (b). 
Inserting Eq.~(\ref{eq:u}) in the detailed-balance condition, one finds that  $p_\infty(\{m\})$ takes a factorized form
\begin{align}
p_\infty(\{m\}) &= {1\over Z_{N,M}}\prod_{i=1}^N f_b(m_i), \nonumber\\
f_b(m)&={1\over \zeta(b) m^b},\nonumber\\
Z_{N,M}&=\prod_{i=1}^N \left(\sum_{m_i=1}^\infty f_b(m_i) \right) \delta\left(\sum_{j=1}^N m_j -M\right),
\label{eq:factor}
\end{align}
with $\zeta(x)$ the Riemann zeta function and $Z_{N,M}$ the partition function. While many properties of the factorized states in Eq.~(\ref{eq:factor}) have been investigated~\cite{PhysRevE.65.026102,PhysRevLett.94.180601,Evans:2006uq}, we have seen  that the case of $b=2$ is the model for the kinetics of product share in international trade, the properties of which are not fully understood.
 
Of main interest is  the particle-number distribution at a single site, which is obtained from the  whole particle configuration probability $p_\infty(\{m\})$ as $p_\infty (m) = \sum_{\{m_2, m_2, \ldots, \}} p_\infty (\{m,m_2,m_3,\ldots\})$ by using Eq.~(\ref{eq:factor})~\cite{0305-4470-38-19-R01}
\begin{align}
p_\infty (m) &=   f_b(m) {Z_{N-1,M-m}\over Z_{N,M}} = {1\over \zeta(b) m^b } {Z_{N-1,M-m}\over Z_{N,M}}.
\label{eq:pm}
\end{align}
The particle-number distribution $p_\infty(m)$ behaves as $p_\infty(m)\simeq f_b(m)=m^{-b}/\zeta(b)$  for $m$ small while the contribution of $Z_{N-1,M-m}/Z_{N,M}$ in Eq.~(\ref{eq:pm}) may not be negligible for $m$ large. When the particle density $\rho$ is small, the latter contribution takes the form $e^{-\mu\,m}$ with   $\mu$ depending on $\rho$ via  the relation $\sum_{m=1}^\infty m \, p_\infty(m) = \rho$. $\mu$ corresponds to the negative chemical potential  $\mu=(\partial/\partial M) \log Z_{N,M}$~\cite{Evans:2006uq}, and decreases with increasing $\rho$. For $b>2$,  there exists the critical density $\rho_c$ such that the relation $\sum_{m=1}^\infty m \, p_\infty(m) = \rho$ can be satisfied by $p_\infty(m) = f_b(m) e^{-\mu\, m}$ with $\mu\geq 0$ only for $\rho\leq \rho_c$:
\begin{equation}
\rho_c = 
{\zeta(b-1)\over \zeta(b)}.
\label{eq:rhoc}
\end{equation}
If $\rho>\rho_c$, $p_\infty(m)$ develops additionally  a bump in the region $m\simeq M-N\rho_c$ 
indicating that most particles occupy a single site, which may be called condensation~\cite{Evans:2006uq}.  For $b\leq 2$ and $\rho$ finite, $p_\infty(m)$ is free from such a bump in the limit $N\to\infty$. Therefore two phases, fluid and condensate phase, can be defined depending on whether such condensation occurs or not in the stationary state, which leads to the phase diagram in the $(b,\rho)$ plane [Fig.~\ref{fig:model}~(b)].

\section{Urn model with $b=2$ for product share kinetics}
\label{sec:modelb2}

The phase diagram  in Fig.~\ref{fig:model} (b) indicates that $b=2$ is a critical value: Condensate  emerges  for finite particle density  only when  $b$ is larger than $2$.  Given that the kinetics of product share in international trade is described by the model with $b=2$,  one can expect that there will be no condensation in international trade. However  in reality $N$ is large but finite and then a bump may appear in $p_\infty(m)$ even with $b\leq 2$ for $\rho$ sufficiently large, which is called pseudo-condensate~\cite{Evans:2006uq}. In this section, we bring the urn model with $b=2$ as close as possible to the real trade by fitting the model parameters, mainly the particle density $\rho$ and the time scale, to address  the model's power of explaining the international trade in the past and predicting its future. 

\subsection{Time scale corresponding to one year}
Let  $\Delta \tilde{t}$ denote the time interval in the model corresponding to one year in the real world. For the time interval $\Delta \tilde{t}$,  a site $p$ is selected $2m_p^2 \Delta \tilde{t}$ times for gaining or losing a particle with the rate in Eq.~(\ref{eq:u}) and $b=2$, which either increases or decreases $m_p$  by~\cite{hughesbook}
\begin{equation}
|\langle \Delta m_p\rangle| \simeq \sqrt{(2/\pi)\mathcal{N}}  \simeq \sqrt{4\Delta \tilde{t}\over \pi} m_p.
\label{eq:Dmpmp}
\end{equation} 
Identifying Eq.~(\ref{eq:DApAp}) with (\ref{eq:Dmpmp}) under the relation   $m_p = M \, A_p$, we find that  $c \simeq \sqrt{4\Delta \tilde{t}\over \pi}$ yielding $\Delta \tilde{t}\simeq 0.01$.  We use the rescaled time  
\begin{equation}
\tau = {\tilde{t}\over \Delta\tilde{t}}
\end{equation}
 such that the increase of $\tau$ by $1$  corresponds to one year in reality. 

\subsection{Fitting the simulated second moment to data}

\begin{figure}
\centering
\includegraphics[width=\columnwidth]{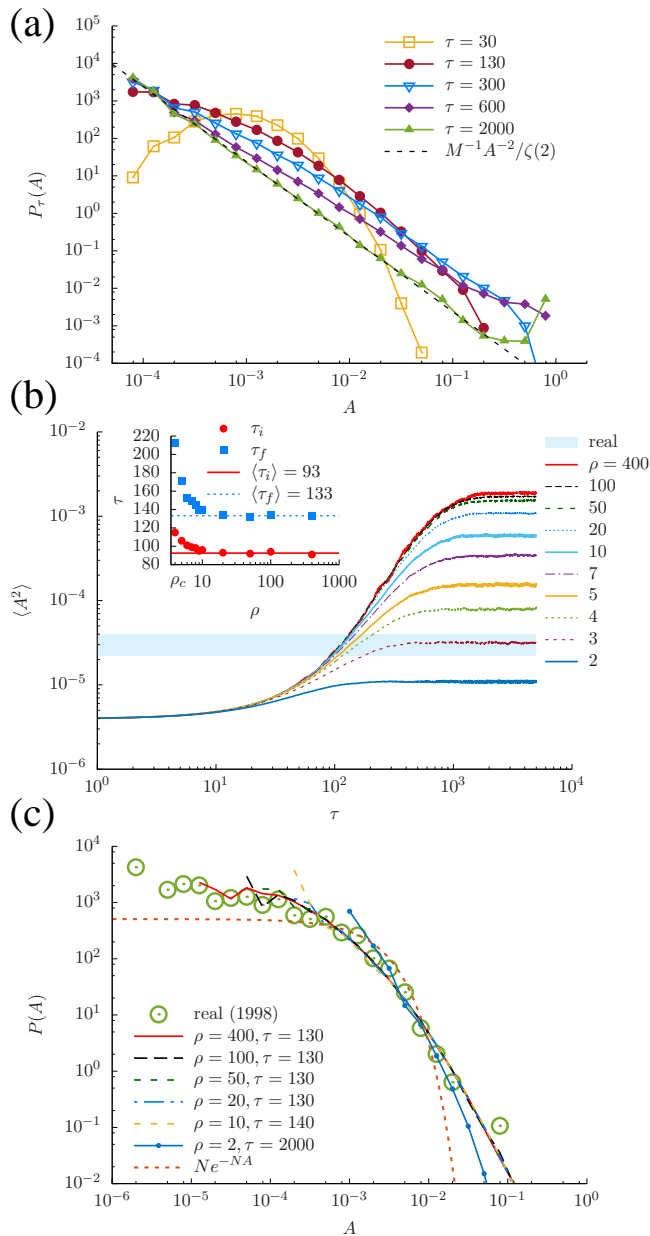}
\caption{Comparison of simulation results and  empirical data. 
(a) The  share distribution at different rescaled times $\tau$ obtained by $P_\tau(A) = M\, p_\tau(m=A\, M)$ from the simulations of the urn model with $b=2$,  the number of sites $N=508$, and the particle density $\rho=50$. The dashed line represents the power law decay.  
(b) The second moment $\langle A^2\rangle$ versus the rescaled time $\tau$ for  different  $\rho$'s. The shaded region indicates the empirical-data range  $2.2\times 10^{-5} \leq \langle A^2\rangle \leq 4.0\times 10^{-5}$. Inset: The rescaled time entering and leaving the empirical-data range $\tau_i$ and $\tau_f$ are given for each value of $\rho$. 
(c) The share distribution $P(A)$ in year 1998 and from model simulations at $\tau=\tau_f$ for each $\rho$. The Boltzmann distribution  $P(A)=N e^{-N\, A}$ with $N=508$ is also shown.
}
\label{fig:density}
\end{figure}

The comparison of the particle-number distribution $p_{\tildet}(m)$ in the model, varying with the particle density $\rho$, and the empirical share distribution may help us to estimate $\rho$. We perform simulations of the model with $N=508$ sites and different particle densities set to be integers. Particles are uniformly distributed over all sites initially at $\tilde{t}=0$, yielding $p_{\tilde{t}=0}(m) = \delta_{m,\rho}$. As the stochastic transfers of particles are repeated, the particle-number distribution  gets broadened as shown in Fig.~\ref{fig:density} (a). The second moment $\langle A^2 \rangle = M^{-2} \langle m^2\rangle = M^{-2}\sum_m m^2 p_{\tau}(m)$ increases with the rescaled time $\tau$ and saturates in the late-time regime as shown in Fig.~\ref{fig:density} (b). The empirical value of $\langle A^2\rangle$ increases steadily from $2\times 10^{-5}$ to $4\times 10^{-5}$ for the period 1962-2000 except for the jumps related to oil crises in the middle of the period, which disappear if crude oil (SITC 3330) is excluded [Fig.~\ref{fig:empirical} (a)]. Interestingly,  the simulated values of $\langle A^2\rangle$ overlap with the empirical values at $90 \leq \tau\leq 130$ as long as $\rho\gtrsim 10$ [Fig.~\ref{fig:density} (b)]. This time interval $\Delta \tau \simeq 40$  is in quite good agreement  with the real time interval, 38 years from 1962 to 2000. For $\rho\lesssim 10$, the overlap period varies with $\rho$  or does not even exist e.g., for $\rho=2$ or $3$. Based on these results, we estimate the particle density  to be $\rho\gtrsim 10$. 

\subsection{Share distribution and condensation}
For all $\rho\gtrsim 10$, the simulated share distributions, $P_\tau(A) =  p_\tau(m=A\, M) M$ in the period $90\leq \tau\leq 130$    match excellently the empirical share distributions as shown in Fig.~\ref{fig:density} (c).   The simulated $P_\tau(A)$ with $\rho=2$  decays faster, failing to fit into the empirical distributions even in the long-time limit. 

Given such excellent agreement between the empirical data and the simulation results in the period $90\leq \tau\leq 130$, one may wonder what will happen in the late-time regime in the simulations.  It can hint at the future of international trade. While the broadening of the simulated distribution $P_\tau(A)$  in the early-time regime does not depend on  the particle density, $P_\tau(A)$ in the late-time regime varies significantly with $\rho$.  The analytic solution for $p_\infty(m)$ in the stationary state, derived in Appendix~\ref{sec:pm}, shows that similarly to $b>2$, a critical density $\rho_c\sim \log N$ exists for $b=2$ and $N$ large but finite such that $Z_{N-1,M-m}/Z_{N,M}$ in Eq.~(\ref{eq:pm})  behaves  as a function of $m$ differently depending on whether $\rho$ is larger than $\rho_c$ or not.  Consequently, for $\rho\lesssim \rho_c$,  $p_\infty(m)$ has an exponential-decay factor 
\begin{equation}
p_\infty(m) \simeq {1\over \zeta(2) m^2} e^{-\mu\, m}
\label{eq:pmasym1}
\end{equation}
with $\mu = e^{-\zeta(2) \rho}>0$. In contrast,  for $\rho\gg \rho_c$, it  has a bump, in addition to the power-law decay form small $m$, as 
\begin{equation}
p_\infty\left(m = M - {N\over \zeta(2)} \log {2N\over\zeta(2)} + {N\over \zeta(2)} y\right) \simeq 
{\zeta(2)\over N^2}  {1\over 2\sqrt{\pi e}} e^{ -{1\over 4} y^2 }
\label{eq:pmasym2}
\end{equation}
for $y$ finite, indicating the emergence of a pseudo-condensate~\cite{Evans:2006uq}: Most particles gather at a single site.  Given $N=508$,  the estimated particle density $\rho\gtrsim 10$  is above the critical density $\rho_c\simeq 3.5$ [Appendix ~\ref{sec:pm}] suggesting that the kinetics of product share will enter the pseudo-condensate phase in the long-time limit: The share distribution is expected to get broadened and end up with a power-law decay plus  a bump near $A=1$ as an example in Fig.~\ref{fig:density} (a) for $\rho=50$. 

The time scale for such condensation phenomena can be predicted by our simulation results. $\langle A^2\rangle$ saturates around $\tau\simeq 10^3$ for $\rho\gtrsim 10$ [Fig.~\ref{fig:density} (b)], which suggest that the stationary-state distribution displaying a bump will appear in $10^3$ years.

The predictions of the capital condensation in the product space and its time scale demonstrate that our model can be a framework for analyzing the evolution of the share distribution at present and in the future. Its limitations  is, however,  worthy to note.  Most of all, we considered a fixed number, 508, of products traded consistently over the studied period 1962-2000. But in reality,  new and old products may enter and leave the space of products, which can make big effects. For instance, the critical density $\rho_c$ will be larger with larger $N$, possibly occurring by the rise of new industries. It might pull the international trade out of the pseudo-condensate phase. Also, annual gains $\Delta A$ show large fluctuations for products having small share $A$.  Allowing larger chance of gain to small-share, probably new, products might help prevent the condensation of share onto top popular products. 

\section{Probabilistic prediction for individual products}
\label{sec:individual}

\begin{figure}
\centering
\includegraphics[width=\columnwidth]{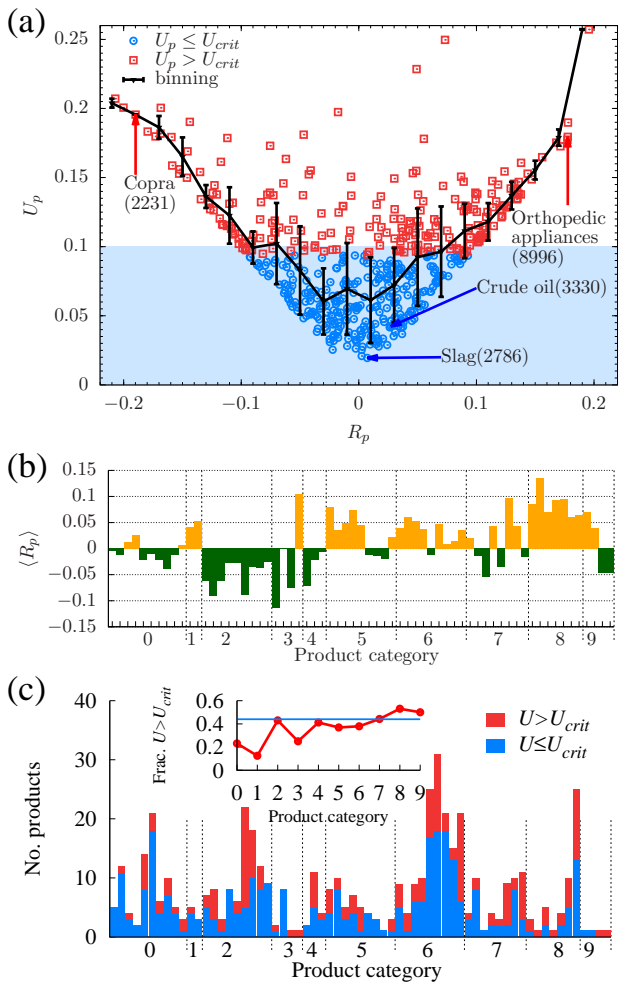}
\caption{Predictability of  individual products' shares. 
(a) Scatter plots of unpredictability $U_p$ and  mean excess growth rate $\langle R_p\rangle=T^{-1}\sum_t R_p(t)$. Blue and red color are used to represent predictable and unpredictable products with the threshold for significance $U_{\rm crit}=0.094$.
(b) The mean excess growth rates of 66 product categories, aggregated by the first 2 digits of their SITC codes. Yellow and green represent whether $\langle R_p\rangle>0$ or $\langle R_p\rangle<0$. 
(c) The number of predictable and unpredictable products in each of the 2-digit categories. 
Inset: The fraction of unpredictable products in each 1-digit SITC category.
}
\label{fig:application}
\end{figure}

Our model can make a probabilistic prediction also for individual products' share. We investigate how well the model can predict the growth rates of share $\{r_p(t)=\log_{10}[A_p(t+1)/A_p(t)]|1962\leq t<2000\}$ for each product $p$. We choose $r_p(t)$'s rather than the share  $A_p(t)$'s since an exceptional variation of the growth rate at a certain time $t_1$ may cause a cascade of shifts in the share $A_p(t)$ for $t\geq t_1$ like the upper shift of share for oil  in Fig.~\ref{fig:empirical} (b). Running simulations with the empirical values at $t=1962$, $\{m_p= M \times A_p(1962)\}$, as the initial configuration at $\tau=0$, we obtain simulated growth rates $\{\tilde{r}_p^{(i)}(\tau)|i=1,2,\ldots, N_{\rm s}\}$ for all products $p$ over the rescaled  time period $0\leq \tau<T=38$ with $\tau = t-1962$ and $N_{\rm s}$ the number of simulation runs. 

From the rank of the empirical growth rate $r_p(t)$ among its $N_{\rm s}$ simulated values $\tilde{r}_p(\tau  = t-1962)$'s, we compute the \textit{excess growth rate}  $R_p(t)$ 
\begin{equation}
R_p(t) = {{\rm rank}\  {\rm of} \ r_p(t) \over N_{\rm s} +2} - {1\over 2},
\label{eq:excess}
\end{equation}
which represents how much the share of product $p$ grows faster (slower) than the median of the simulated growth rates. The $T$ excess growth rates $\{R_p(t)|1962\leq t\leq 2000\}$ for a product $p$ will be uniformly distributed between $-1/2$ and $1/2$ if simulations yield a good probabilistic prediction for the empirical growth rates $r_p(t)$'s of the product $p$~\cite{marhuenda2005comparison}; it should not be possible to discern empirical values  from the simulated ones. 
Let us sort $t$'s such that  $R_p(t_1)\leq R_p(t_2)\leq \cdots \leq R_p(t_T)$. If $R_p(t)$'s are uniformly distributed between $-1/2$ and $1/2$, then we expect that $R_p(t_i)$'s will increase linearly with $i$ such that $R_p(t_i) = \bar{R}_i = i/(T+1) -1/2$. 
Deviations from this expectation can quantify the non-uniformity of $R_p(t)$'s and the failure of the model to predict the growth rates $r_p(t)$' of product $p$~\cite{marhuenda2005comparison}. 
Therefore we define the unpredictability $U_p$ of product $p$  by 
\begin{equation}
U_p=T^{-1} \sum_{i=1}^T |R_p(t_i) - \bar{R}_i|,
\label{eq:Up}
\end{equation}  
where the $t_i$'s are sorted such that $R_p(t_i)\leq R_p(t_j)$ for all $i<j$ and $\bar{R}_i=\frac{i }{ T+1}-\frac{1}{ 2}$. $U_p$ of $T=38$  numbers  from the uniform distribution  may be larger than $0.094$ with probability $0.05$. Therefore we consider products with  $U_p>U_{\rm crit} = 0.094$ as unpredictable~\cite{marhuenda2005comparison}. 

We find that 284 products ($56\%$) are predictable  while the remaining 224 ($44\%$) are unpredictable by the model simulations with  $\rho=400$. The number of predictable products was smaller for other selected values of $\rho$. More products are classified as predictable  if the model prediction is made for a shorter period than $T=38$ years. Most of the unpredictable products display large positive or negative excess growth rate (Fig.~\ref{fig:application} (a)). Interestingly, such deviations are correlated with the nature of the products, as identified in the SITC framework: Growth rates for raw materials and agricultural commodities tend to be smaller than the predicted growth rates, showing the mean excess growth rate  $\langle R_p\rangle=T^{-1}\sum_t R_p(t)$ negative, while manufactured and especially high-end products such as medical appliances have  $\langle R_p\rangle$ positive. Figure~\ref{fig:application} (b) shows the mean excess growth rates of  66 product categories based  on the first two digits  of  SITC. Chemicals (SITC codes beginning with 5), manufactured goods (6 and 8) and machinery (7) have grown faster than the model prediction   while  the growth of the share of crude materials (2) was slower than the prediction.  The fraction of unpredictable products is larger among manufactured goods than among raw materials (Fig.~\ref{fig:application} (c)).

Some products show small $|R_p|$ but large $U_p$ owing to large fluctuations $\sigma_p^2 = T^{-1}\sum_t (R_p(t)-\langle R_p\rangle)^2$; the offer and demand for such products tends to be highly variable or historically determined e.g. railways, warships, and uranium. Thus, our predictability metric captures the fact that some products are more (or less) variable than expected, even if their trend is  predicted.

\section{Summary and discussion}

In this paper we presented a stochastic particle model for the kinetics of products' shares in international trade, in which sites represent product categories and particles represent a unit of share. From the empirical  scaling behavior of the annual variations of product share,   the probability of a site to send a particle to another site is set to be proportional to the square of the number of particles at the two sites. Products' shares are related to the capital invested in those product categories,  and therefore such biquadratic transfer rate illuminates a fundamental nature of capital in its activity and interaction.

If the preference to popular products  is weaker than that represented by the biquadratic rate, the stationary state will lie in the fluid phase, having no condensation as long as the particle density is finite. Therefore  capital in international trade may be considered as being at the edge of the fluid phase.  Comparing with the empirical data,  we found that the period 1962-2000 covered in the dataset corresponds to  a transient period in the model. We were able to determine the time scale and the range of the particle density with which the model predictions match the empirical annual variation and distribution of product share. The estimated particle density turns out to be larger than the critical density, suggesting that a large fraction of  trade volumes will concentrate on few product categories in the future. The model simulations show how the share distribution will change with time and predict the time scale of such condensation by using the estimated parameters. The model also makes baseline predictions for the evolution of various industrial sectors, and by comparison, allows us to find other factors such as intrinsic economic fitness~\cite{safarzynska2010evolutionary} at work in addition to the endogenous dynamics represented by our model.

While our model can serve as a simple framework for analyzing the kinetics of capital in the space of products, it could be extended by allowing for the birth and death of product categories, leading to expansion or contraction of the product space. Also  intrinsic fitness of products can be considered. Another natural development would be to assume that capital transfers occur  on a network of products linked through heterogeneously weighted development pathways~\cite{hidalgo2007product,hidalgo2009building,hausmann2011network}.

\begin{acknowledgements}
This work was supported by the National Research Foundation of Korea (NRF) grants funded by the Korean Government (No. 2013R1A1A201068845 and No. 2016R1A2B4013204).
\end{acknowledgements}

\appendix

\section{Derivation of the single-site distribution in the stationary state of the urn model with $b=2$}
\label{sec:pm}

In Eq.~(\ref{eq:factor}), the factorized form of the configuration probability $p_\infty(\{m\}) = Z_{N,M}^{-1} \prod_i f_2 (m_i)$ with $f_2(m) = m^{-2}/\zeta(2)$ leads us to find that the grand partition function is given by $\Zgpf_N (z) = \sum_{M=0}^\infty z^M Z_{N,M}= F_2(z)^N$ with $F_2(z)$ the generating function  of $f_2(m)$ represented as   
$F_2(z)\equiv\sum_{m=1}^\infty  z^m f_2(m) = \sum_{m=1}^\infty {z^m \over \zeta(2) m^2} = {{\rm Li}_2 (z) \over \zeta(2)}$, where ${\rm Li}_b(z)$ is the poly-logarithm function ${\rm Li}_b(z) = \sum_{m=1}^\infty z^m m^{-b}$. This series converges for $|z|\leq 1$ and 
expanded around $z=1$ as~\cite{PhysRev.83.678}
\begin{equation}
F(z=e^{-\alpha}) = 1 + {\alpha \log \alpha -\alpha  \over \zeta(2)} + \sum_{n=2}^\infty {(-1)^n \over n!} {\zeta(2-n) \over \zeta(2)} \alpha^n.
\label{eq:Fz}
\end{equation}

The partition function $Z_{N,M}$ can be recovered from the grand partition function by $Z_{N,M}=\oint {dz \over 2\pi i} z^{-M-1} F(z)^N$, where the contour is within the radius of convergence, $|z|\leq 1$~\cite{0305-4470-38-19-R01}. Using Eq.~(\ref{eq:Fz}), one finds that  the dominant contribution  is made around $z_*=e^{-\mu}$ with 
\begin{equation}
\mu = e^{-\rho \zeta(2)},
\end{equation}
 and thus the integral  is approximated in the limit $\mu\to 0$, $N\to\infty$, and $N\mu$ finite by employing the steepest descent path $z=e^{-\mu y}$ with $y$ running from $1-i\infty$ to $1+i\infty$ as 
\begin{equation}
Z_{N,M}=\mu\, \phi\left({N\mu \over \zeta(2)}\right), \ \phi(\eta)=\int_{1-i\infty}^{1+i\infty} {dy \over 2\pi i} e^{\eta (y \log y - y)}.
\label{eq:ZNM}
\end{equation}
Using Eq.~(\ref{eq:ZNM}) into Eq.~(\ref{eq:pm}), we find that 
\begin{equation}
p_\infty(m) = {1\over \zeta(2) m^2} e^{{m\over N}\zeta(2)} {\phi\left({N\mu \over \zeta(2)}e^{{m\over N}\zeta(2)}\right)\over \phi\left({N\mu \over \zeta(2)}\right)}.
\label{eq:pmexact}
\end{equation}

For $\phi(\eta)$ defined in Eq.~(\ref{eq:ZNM}), when $\eta$ is large, the contribution  near $y=0$ to $\phi(\eta)$ is dominant, allowing us to approximate it as
\begin{equation}
\phi(\eta) \simeq \int_{-\infty}^\infty {d\beta \over 2\pi} e^{-{\eta} ( 1 + {\beta^2 \over 2})} \simeq  \sqrt{1\over 2\pi \eta} e^{-{\eta}}.
\label{eq:phi1}
\end{equation}
According to our numerical computation, this approximation is good even for $\eta\simeq 1$. For $\eta\ll 1$, we consider the path surrounding the branch cut of $\log y$, $y=x \pm i\epsilon$ with $x\in (-\infty,0)$ and $\epsilon\to 0$  such that 
$\phi(\eta)= \int_{-\infty}^0 {dx \over 2\pi i} e^{-\eta ( x \log (x-i\epsilon) -  x)} + \int_0^{-\infty} {dx \over 2\pi i} e^{-\eta ( x \log (x+i\epsilon) -  x)}=\int_0^{\infty} {dx \over \pi} e^{\eta ( x - x \log x)} \sin (\eta \pi x)$, which leads us to
\begin{align}
\phi(\eta) &= \int_0^{\infty} {dq \over \eta \pi} e^{ ( q - q \log q + q \log \eta)} \sin (\pi q) \nonumber\\
 &\simeq \int_0^{\infty} {dq \over \eta} q\, e^{  q \log \eta}   \simeq {1\over \eta |\log \eta|^2}.
\label{eq:phi2}
\end{align}

\begin{figure}
\centering
\includegraphics[width=\columnwidth]{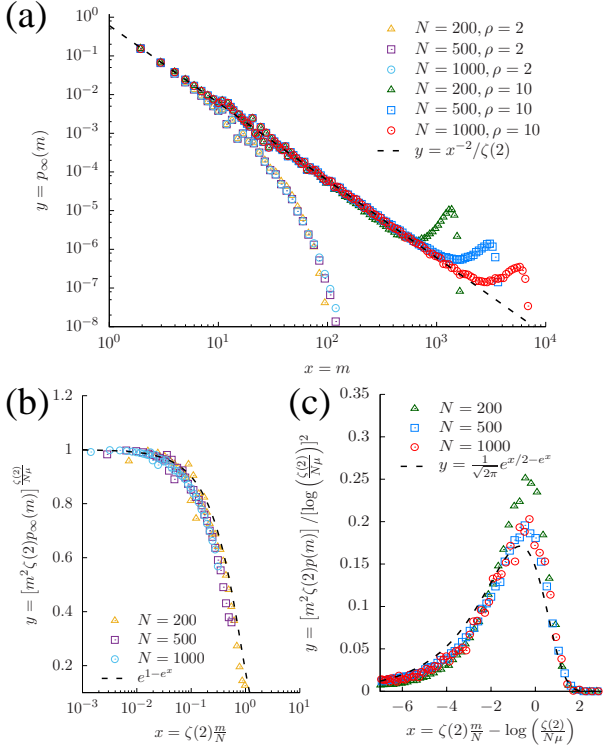}
\caption{The  particle-number distribution $p_\infty(m)$ in the urn model with $b=2$. (a) $p_\infty(m)$ from simulations decays exponentially for $\rho=2$ while it exhibits a bump for $\rho=10$  at different locations depending on the system size $N$. The dashed line indicates the power-law decay $f_2(m)$ in Eq.~(\ref{eq:pm}). (b) Data collapse of $[m^2 \zeta(2) p_\infty(m)]^{\zeta(2)/N\mu}$ versus $\zeta(2) m/N$    for different $N$'s and $\rho=2$ (fluid phase). The dashed line represents the theoretical prediction $y=e^{1-e^x}$ in Eq.~(\ref{eq:fluid}). (c) Data collapse of bumps, $[m^2 \zeta(2) p_\infty(m)]/[\log (\zeta(2)/(N\mu))]^2$ versus $\zeta(2)m/N-\log(\zeta(2)/(N\mu))$ for different $N$'s and $\rho=10$ (condensate phase). The dashed line is $y = {1\over \sqrt{2\pi}} e^{x/2 - e^x}$ from Eq.~(\ref{eq:condensate}). 
}
\label{fig:stationary}
\end{figure}

Different  behaviors of $\phi(\eta)$ for $\eta \gtrsim 1$ and $\eta\ll 1$ in Eqs.~(\ref{eq:phi1}) and (\ref{eq:phi2}) give rise to  different behaviors of $p_\infty(m)$ depending on $\rho$ and $N$ as shown in Fig.~\ref{fig:stationary} (a). In case $N\mu/\zeta(2)\gtrsim 1$, corresponding to the low-density regime $\rho\lesssim \rho_c$ with the critical density defined by
\begin{equation}
 \rho_c \equiv {1\over \zeta(2)} \log {N\over \zeta(2)},
\label{eq:rhocb2}
\end{equation}
both $\phi$ functions in Eq.~(\ref{eq:pmexact}) behave as Eq.~(\ref{eq:phi1}). Therefore the particle-number distribution behaves as 
\begin{align}
p_\infty(m) &\simeq {1\over \zeta(2) m^2} e^{-{N\mu \over \zeta(2)} (e^{{m\over N}\zeta(2)}-1)},
\label{eq:fluid}
\end{align}
which is confirmed by the simulation results for various $N$'s and $\rho=2$ [Fig.~\ref{fig:stationary} (b)]. One finds that in the regime $x = {N\over \zeta(2)} m \ll 1$,  $p_\infty(m)$  has the exponential-decaying term  as in Eq.~(\ref{eq:pmasym1}).

If $N\mu/\zeta(2)\ll 1$ or $\rho\gg \rho_c$, the $\phi$ function in the denominator in Eq.~(\ref{eq:pmexact}) behaves as Eq.~(\ref{eq:phi2}). In the regime ${\zeta(2) \over N} [m - (M-M_c)] \ll -1$ with $M_c= N \rho_c = [N/\zeta(2)] \log[N/\zeta(2)]$, the $\phi$ function in the numerator in Eq.~(\ref{eq:pmexact}) behaves also as Eq.~(\ref{eq:phi2}) and we find that the ratio of the two $\phi$ functions is close to $1$, leading to $p_\infty(m) \simeq {1\over \zeta(2) m^2}$.  In the large-$m$ region where ${\zeta(2) \over N} [m - (M-M_c)] \gtrsim 1$, the $\phi$ function in the numerator in Eq.~(\ref{eq:pmexact}) behaves  as Eq.~(\ref{eq:phi1}), leading to 
\begin{equation}
p_\infty(m) = {1\over \zeta(2) m^2} \left[\log\left({\zeta(2) \over N\mu}\right)\right]^2 {1\over \sqrt{2\pi}} e^{{x\over 2} - e^x},
\label{eq:condensate}
\end{equation} 
with $x= {\zeta(2) \over N} [m - (M-M_c)]=\zeta(2)({m\over N} - \rho  + \rho_c) = \zeta(2) {m\over N} - \log\left({\zeta(2) \over N\mu} \right)$. The function $e^{x/2 - e^x}$ takes a bell-shape around $x_0 = -\log 2$ such that 
it takes a form ${1\over \sqrt{2e}} e^{-{1\over 4} (x-x_0)^2}$ around $x_0$ which leads to Eq.~(\ref{eq:pmasym2}). The scaled plots of $p_\infty(m)$ in the regime where $x$ is finite in Fig.~\ref{fig:stationary} (c) confirm Eq.~(\ref{eq:condensate}).

Using Eqs.~(\ref{eq:pmexact}), (\ref{eq:fluid}) and (\ref{eq:condensate}), we can represent $p_\infty(m)$ as
\begin{equation}
p_\infty(m) = {1\over \zeta(2) \, m^2} \Psi\left(\zeta(2){m\over N}; \zeta(2)(\rho_c-\rho)\right),
\end{equation}
where  the behaviors of $\Psi(\theta;\lambda) \equiv e^\theta {\phi\left(e^{\theta+\lambda}\right)\over \phi\left(e^\lambda\right)}$ can be summarized as
\begin{equation}
\Psi(\theta;\lambda) \simeq 
\left\{
\begin{array}{ll}
e^{-e^\lambda (e^\theta -1)} & {\rm for} \, \lambda\gtrsim 1, \\
1 & {\rm for} \, \lambda\ll -1,  \theta+\lambda\ll -1,\\
{\lambda^2 \over \sqrt{2\pi}} e^{{1\over 2} (\theta+\lambda) - e^{\theta+\lambda}} & {\rm for} \, \lambda\ll -1, \theta+\lambda \gtrsim 1.
\end{array}
\right.
\end{equation}
The critical density $\rho_c$ is about $3.5$ for $N=508$ according to Eq.~(\ref{eq:rhocb2}). 

%

\end{document}